\documentclass{acm_proc_article-sp}
\usepackage{stackengine}[2013-10-15]
\usepackage{fixltx2e}
\usepackage{color}
\begin{document}

\title{Precision-Aware application execution  for Energy-optimization in HPC node system
}

%
%
%
%
%

\numberofauthors{8} 
%
\author{
%
%
\alignauthor
Radim Vavrik \\
       \affaddr{IT4Innovations National Supercomputing Center}\\
       \affaddr{VSB-Technical University of Ostrava}\\
       \affaddr{Ostrava, Czech Republic}\\
       \email{radim.vavrik@vsb.cz}
\alignauthor
Antoni Portero\\
       \affaddr{IT4Innovations National Supercomputing Center}\\
       \affaddr{VSB-Technical University of Ostrava}\\
       \affaddr{Ostrava, Czech Republic}\\
       \email{antonio.portero@vsb.cz}
\alignauthor 
Stepan Kuchar \\
       \affaddr{IT4Innovations National Supercomputing Center}\\
       \affaddr{VSB-Technical University of Ostrava}\\
       \affaddr{Ostrava, Czech Republic}\\
       \email{stepan.kuchar@vsb.cz}
\and  
\alignauthor 
Martin Golasowski\\
       \affaddr{IT4Innovations National Supercomputing Center}\\
       \affaddr{VSB-Technical University of Ostrava}\\
       \affaddr{Ostrava, Czech Republic}\\
       \email{martin.golasowski@vsb.cz}
\alignauthor 
Simone Libutti \\
       \affaddr{Politecnico di Milano}\\
       \affaddr{Dipartimento di Elettronica, Informazione e Biongegneria (DEIB)}\\
       \affaddr{ Milano, Italy}\\
       \email{simone.libutti@polimi.it}
\alignauthor 
Giuseppe Massari\\
       \affaddr{Politecnico di Milano}\\
       \affaddr{Dipartimento di Elettronica, Informazione e Biongegneria (DEIB)}\\
       \affaddr{ Milano, Italy}\\
       \email{giuseppe.massari@polimi.it}
}

\additionalauthors{Additional authors: William Fornaciari (Politecnico di Milano,
email: {\texttt{william.fornaciari@polimi.it}}) and Vit Vondrak
(IT4 Innovations National Supercomputing Center, email: {\texttt{\\vit.vondrak@vsb.cz}}).}
\date{30 July 1999}

\maketitle



\begin{abstract}
Power consumption is a critical consideration in high performance computing systems and it is becoming the limiting factor to build and operate Petascale and Exascale systems. When studying the power consumption of existing systems running HPC workloads, we find that power, energy and performance are closely related which leads to the possibility to optimize energy consumption without sacrificing (much or at all) the performance. In this paper, we propose a HPC system running with a GNU/Linux OS and a Real Time Resource Manager (RTRM) that is aware and monitors the healthy of the platform. On the system, an application for disaster management runs. The application can  run with different QoS depending on the situation. We defined two main situations. Normal execution, when there is no risk of a disaster, even though we still have to run the system to look ahead in the near future if the situation changes suddenly. In the second scenario, the possibilities for a disaster are very high. Then the allocation of more resources for improving the precision and the human decision has to be taken into account. The paper shows that at design time, it is possible to describe different optimal points that are going to be used at runtime by the RTRM with the application. This environment helps to the system that must run 24/7 in saving energy with the trade-off of losing precision. The paper shows a model execution which can improve the precision of results by 65\% in average by increasing the number of iterations from 10\textsuperscript{3} to 10\textsuperscript{4}. This also produces one order of magnitude longer execution time which leads to the need to use a multi-node solution. The optimal trade-off between precision vs. execution time is computed by the RTRM with the time overhead less than 10\% against a native execution.
\end{abstract}

\category{}{HPC systems, disaster management, reliability models, computing resource management, many-cores, parallel systems}[PACS: 89.20.Ff, 47.50.Cd, 92.40.-t, 07.05.Bx, 07.05.Tp, 89.20.Ff]
 


\section{Introduction}

Application requirements, power, and technological constrain\-ts are driving the architectural convergence of future processors towards heterogeneous many-cores. This development is confronted with variability challenges, mainly the susceptibility to time-dependent variations in silicon devices. Increasing guard-bands to battle variations is not scalable, due to too large worst-case cost impact for technology nodes around 10 nm. The goal of next generation firmware is to enable next-generation embedded and high-performance heterogeneous many-cores to cost-effectively confront variations by providing Dependable-Performance: correct functionality and timing guarantees throughout the expected lifetime of a platform under thermal, power, and energy constraints.
An optimal solution should employ a cross-layer approach. A middle-ware implements a control engine that steers software/hardware knobs based on information from strategically dispersed monitors. This engine relies on technology models to identify/exploit various types of platform slack - performance, power/energy, thermal, lifetime, and structural (hardware) - to restore timing guarantees and ensure the expected lifetime and time-dependent variations. 

Dependable-Performance is critical for embedded applications to provide timing correctness; for high-performance applications, it is paramount to ensure load balancing in parallel phases and fast execution of sequential phases. The lifetime requirement has ramifications on the manufacturing process cost and the number of field-returns. The future firmware novelty must rely in seeking synergies in techniques that have been considered virtually exclusively in the embedded or high-performance domains (worst-case guaranteed partly proactive techniques in embedded, and dynamic best-effort reactive techniques in high-performance).
This possible future solutions will demonstrate the benefits of merging concepts from these two domains by evaluating key applications from both segments running on embedded and high-performance platforms. The intent of this paper is to describe the characteristics and the constraints of disaster management (DM) applications for industrial environments. When defining the requirements and their evaluation procedure, a first analysis of the DM applications modules (HW platform, OS and RT engines, monitors and knobs, reliability models) is provided.
Power consumption is a critical consideration in high performance computing systems and it is becoming the limiting factor to build and operate Petascale and Exascale systems. When studying the power consumption of existing systems running HPC workloads, we find power, energy and performance are closely related leading to the possibility  to optimize energy without sacrificing (much or at all) performance. In this paper, we propose a HPC system running with an  RTRM that is aware and monitors the healthy of the platform. On the system, an application for disaster management is running. The application can  run with different QoS depending on the situation. We defined two main situations. Normal execution when there is no risk of disaster, even though we still have to run the system to look ahead in the near future if the situation changes suddenly. And the second scenario where the possibilities for a disaster are very high. Then the allocation of more resources for improving the precision and human decision has to be taken into account. The paper shows that at design time, it is possible to describe different optimal points that are going to be used at runtime by the RTRM with the application. This environment helps to the system that must run 24/7 in saving energy with the trade-off of losing precision.

The paper shows a model execution which can improve the precision of results by 65\% in average by increasing the number of iterations from 10\textsuperscript{3} to 10\textsuperscript{4}. This also produces one order of magnitude longer execution time which leads to the need to use a multi-node solution. The optimal trade-off between precision vs. execution time is computed by the RTRM with the time overhead less than 10\% against a native execution.
 
The paper is divided into nine sections, after the introduction the state-of the art follows. In section three, we introduce our driving disaster management example named the uncertainty modelling of rainfall-runoff model. Section four describes the Real Time Resource Manager (RTRM) that monitors the sensors and knobs of the platform. Section five shows results of possible optimal computation points that are obtained at the design time. They are going to be used by the RTRM at runtime in near future. Finally, there is a section of conclusion and future work. 

\section{The State of the Art}

One of the goals for future HPC systems is to develop automated frameworks that use power and performance to make applications-aware energy optimizations during an execution. Techniques like dynamic voltage and frequency scaling (DVFS) for reducing the speed (clock frequency) in exchange for reduced power consumption are used \cite{Anagnostopoulos13}. And power gating technique that allows reducing power consumption by shutting off the current to blocks of the circuit that are not in use. Different computations have different power requirements. For computations where the CPU is waiting for resources the frequency can be reduced to lower power with minimal performance impact \cite{Tiwari12}.
System scenarios classify system behaviours that are similar from a multidimensional cost perspective, such as resource requirements, delay and energy consumption, in such a way that the system can be configured to exploit this cost. At design-time, these scenarios are individually optimized. Mechanisms for predicting the current scenario at run-time and for switching between scenarios are also derived. These are derived from the combination of the behaviour of the application and the applications mapping on the system platform.
These scenarios are used to reduce the system cost by exploiting information about what can happen at run-time to make better design decisions at design-time, and to exploit the time-varying behaviour at run-time. While use-case scenarios classify the application's behaviour based on the different ways the system can be used in its over-all context, system scenarios classify the behaviour based on the multi-dimensional cost trade-off during the implementation \cite{Gheorgita09}.

To get precise results as fast as possible is the main requirement for our testing application with the risk of disaster scenario. Runing applications with such requirements is called urgent computing \cite{Galante12}. There are plenty of specifics for such kind of computing, e.g. need for infrequent, but massive computational performance (HPC). Since reserving computing capacity only for urgent cases is economically inefficient, the usage of cloud computing can be the solution. On the other hand, there are limitations for communication-intensive applications and guarantee the availability of the resources may be problematic.

\subsection{RTRM: Run-Time Resource Manager }

The framework \cite{Bellasi12},\cite{Harpa13} is the core of a highly modular and extensible run-time resource manager which provides support for an easy integration and management of multiple applications competing on the usage of one (or more) shared MIMD many-core computation devices. The framework design, which exposes different plug in interfaces, provides support for pluggable policies for both resource scheduling and the management of applications coordination and reconfiguration.
Applications integrated with this framework get a suitable instrumentation to support Design-Space-Explorati\-on (DSE) techniques, which could be used to profile application behaviours to either optimize them at design time or support the identification of optimal QoS requirements goals as well as their run-time monitoring. Suitable platform abstraction layers, built on top of GNU/Linux kernel interfaces, allow an easy porting of the framework to different platforms and its integration with specific execution environments. We use a common GNU/Linux with an efficient Run-Time Resource Manager (RTRM) \cite{Bellasi12},\cite{Harpa13} that exploits a Design-Time Exploration (DSE), which performs an optimal quantization of the configuration space of run-time tunable applications, identifying a set of configurations. The configuration of a run-time tunable application is defined by a set of parameters. Some of them could impact the application behaviour (for example the uncertainty of Rainfall-Runoff (RR) models can produce different accuracy for different situations) while other have direct impact on the amount of required resources. For example, the amount of Monte Carlo iterations in uncertainty modelling and the time between batches of simulations lead to different requirements for allocating system resources.

\section{Disaster Management Model}
The model under study is a modular part of the FLOREON+ project \cite{Martinovic10, Theuer14}. The main objective of the FLOREON+ project is to create a platform for integration and operation of monitoring, modelling, prediction and decision support for disaster management. Modularity of the FLOREON+ system, which is developed for this science and research project, allows for simple integration of different thematic areas, regions and data. This system is meant to be used for decision support in operative and predictive disaster management processes. The central thematic area of the project is hydrologic modelling and prediction. The system focuses on acquisition and analysis of relevant data in real time and application of prediction algorithms with this data. The results are then used for decision support in disaster management processes by providing predicted discharge volumes on measuring stations and prediction and visualization of flood lakes in the landscape.
Model in this study \cite{Golasowski14} is used for short term predictions of hydrological situation on given rivers. Rainfall-runoff (RR) model is a dynamic mathematical model which transforms rainfall to flow at the catchment outlet. The main purpose of the model is to describe rainfall-runoff relations of a catchment area. Common inputs of the model are precipitations on given stations and spatial and physical properties of the river catchment area. Common outputs are surface runoff hydro-graphs which depict relations between discharge (Q) and time (t). Catchment areas for the model used in this paper for experiments are parametrized by following values. Manning's roughness coefficent (N) which approximates physical properties of the river basin and CN curve number which approximates geological properties of the river catchment area.
Rainfall-runoff models are usually used for predicting behaviour of river discharge and water level and one of the inputs of rainfall-runoff models is the information about weather conditions in the near future. These data are provided by numerical weather prediction models  \cite{Golasowski14} and can be affected by certain inaccuracy. Such inaccuracy can be projected into output of the model by constructing confidence intervals which provide additional information about possible uncertainty of the model output. The confidence intervals are constructed by using iterative Monte Carlo method. Data sets are sampled from the model input space and used as input for a large number of simulations. Quantiles selected from the simulations output are merged with the original hydrograph. Precision of modelling of the model uncertainty can be positively affected by increasing the number of performed simulations. The precision of the simulations can be determined by estimating the Nash-Sutcliffe (N-S) model efficiency coefficient between the original simulation output and one of the quantiles selected from the Monte Carlo results. The N-S coefficient is often used for estimating precision of given model by comparing its output with measured data but can be successfully used for comparison of any two model outputs.

\begin{equation}E= 1-\frac{\sum_{t=1}^{T}(Q_{0}^{t}-Q_{m}^{t})^2}{\sum_{t=1}^{T}(Q_{0}^{t}-\overline{Q_{0}^{t}})^2} \end{equation}

Where:\newline 
\small{Q\textsubscript{0} is the measured flow in one timestep.}\newline 
\small{Q\textsubscript{m} is the simulated flow in one timestep.}\newline 
\small{E is the Nash-Sutcliffe model efficiency coefficient.}\newline
\small{E = 1 means that the simulation matches observed data perfectly.}\newline 
\small{E = 0 simulation matches median of the observed data.}\newline 
\small{E < 0 simulation is less precise than the median of the observed data.}\newline 

Section \ref{results} presents our study about the precision that the model provides and its relation with the amount of computation. As we explain later, at design time, it is possible to simulate situations with different trade-offs in terms of quality of services (QoS), execution time and power-consumption. Then the RTRM choose the optimal working point depending on the application scenarios.

\subsection{Application Scenarios}

Application scenarios \cite{CISIM14} describe different triggers and states of the application that influence the system responsiveness and operation (e.g. critical flooding level, critical state of patient's health, voice \& data, etc.). Based on these scenarios, the system can be in different states with different service level requirements that can also be translated to the parameters of the uncertainty modelling and its required resources. Shorter response time in critical situations can for example be acquired by decreasing the number of Monte Carlo iterations (also decreasing the precision of the results), or by allocating more computational resources (maintaining the same precision level if the load is rebalanced appropriately). We identified two main application scenarios that support the different workload of the system based on the flood emergency situation.

\subsubsection{Standard Operation}
In this scenario, weather is favourable and the flood warning level is below the critical threshold. The computation can be relaxed; some level of errors and deviations can be allowed. The system should only use as much power as needed for standard operation; one batch of RR simulations with uncertainty modelling only has to be finished before the next batch starts. The results do not have to be available as soon as possible, so no excess use of resources is needed.  
\subsubsection{Emergency Operation}
Several days of continuous rain raise the water in rivers or a very heavy rainfall on a small area creates new free-flowing streams. These conditions are signalled by the discharge volume exceeding the flood emergency thresholds or precipitation amount exceeding the flash flood emergency thresholds. Much more accurate and frequent computations are needed in this scenario and results has to be provided as soon as possible even if excess resources have to be allocated. 

\section{RTRM Methods: Abstract Execution Model}

The development of the runtime reconfigurable application is facilitated using a real time library (RTLIB). RTLIB provides 
the Abstract Execution Model (AEM). The AEM encapsulates the execution flow of the application in a way that let the real time resource manager RTRM \cite{Bellasi12} to govern its life-cycle. Generally, we can structure an application by splitting it in more Execution Contexts (EXC). From the RTRM point of view an EXC is seen as  an schedulable task. Having more EXC in the same application can come from the need of schedule parts of the application with different priorities and resource usages. This is made possible using the recipe that is associated to the Execution Contexts \cite{bosp14}.

\begin{figure}
\centering
\epsfig{file=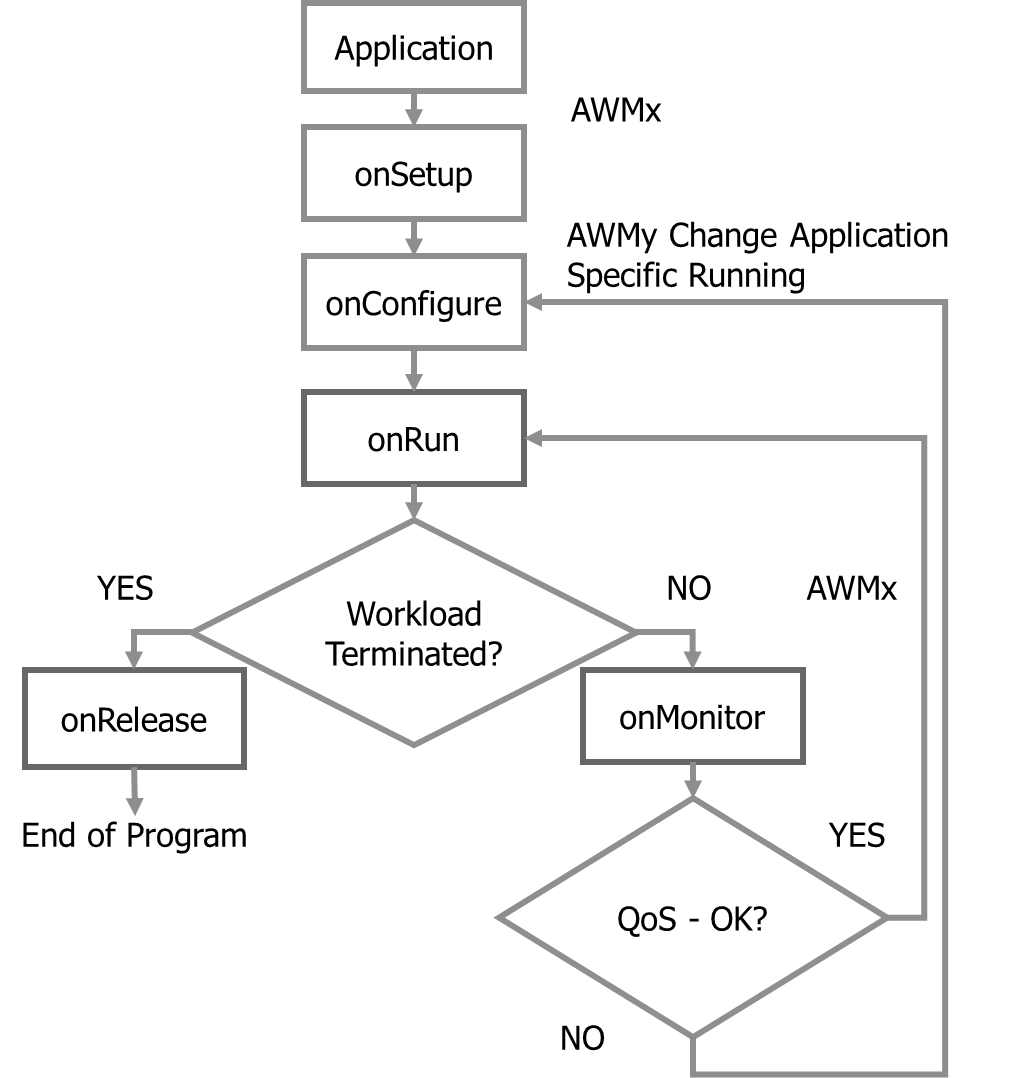, height=3.5in, width=3.0in}
\caption{Diagram of the RTRM methods for uncertainty Application.}
\label{h_flow}
\end{figure}

From the application side, the mandatory requirement to exploit the RTRM is to provide an implementation that allows it to be run-time reconfigurable. From the RTRM side, the configuration of an application is characterized in terms of set of resource requirements, and is called Application Working Mode (AWM). The application must declare to the RTRM the list of its Application Working Modes. Such list must be provided through a file called, in jargon, Recipe which basically is an XML file.  Fig. \ref{h_flow} summarizes the Abstract Execution Model, where the Application, onConfigure, Workload Terminated? and QoS - OK? boxes  are related to the member methods of the derived class. As we can see, the AEM configures a sort of state diagram. Thus the methods must implement what our Execution Context must do whenever reach the corresponding execution state. 

\begin{figure}
\centering
\epsfig{file=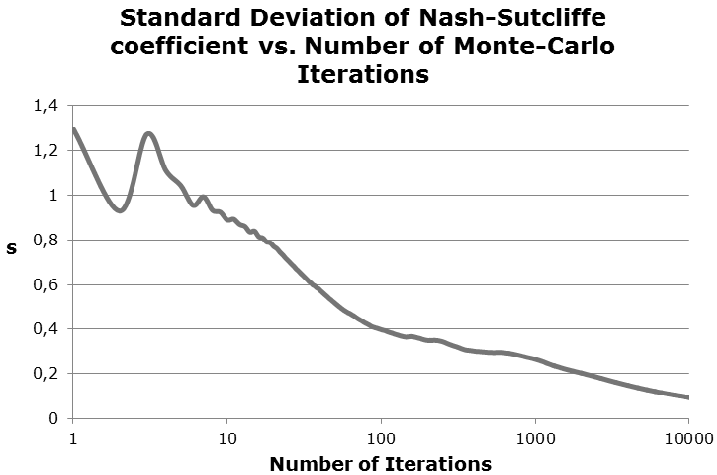, height=2.8in, width=3.5in}
\caption{Precision vs. number of iterations.}
\label{precision}
\end{figure}

\subsection*{Method onSetup}
Initialization of the framework is performed in the \textit{onSetup} function. In order to allow the RTRM to perform a correct initialization of the structures supporting the collection of the runtime statistics, this method performs the initialization of the structure named "val". It has all the input values that can modify the uncertainty simulation. The description of the values is:
\begin{itemize}
\item \textit{CN, N  range and precision}: Probability distribution function and its parameters used for generating samples of the CN number and Manning's rougness coefficient values.
\item \textit{Input Rain Data}: The data about rainfall on the land region under study received from the weather stations. New data is received from there each ten minutes to one hour.
\item \textit{Threads for uncertainty}: The number of (OpenMP) threads equivalent to the number of Monte Carlo executions of the rainfall-runoff model that runs in parallel to simulate the uncertainty.
\end{itemize}
Threads for rainfall-runoff model: The number of (OpenMP) threads that parallelize the main loops of the rainfall-runoff model itself.
\subsection*{Method onConfigure}
Called when the RTRM is assigned for the first time, or the Application Working Mode (AWM) has changed, i.e., the set of resources allocated for the application/EXC. The code of this method is related to whatever is required to reconfigure the application (number of threads, application parameters, data structures, etc.) to properly run according to resources assigned through the AWM.
\subsection*{Method onRun}
This is the entry point of the task. In our case the uncertainty simulation. There is a call to our wrapped 'main' method called \textit{'floreon-model'}. It has a structure ( named \textit{val}) as an input, which provides input parameters that can vary (modify) the performance of the model. 
 
\begin{equation}(Result, Quality) = f[ \textit{floreon-model} (   val)];\end{equation}
Where \begin{equation} Quality=f(E,t); \end{equation} \textit {E-Precision(Nash-Sutcliffe model efficiency coefficient), t-time }

\subsection*{Method onMonitor}
After a computational run, the application may check whether the level of QoS/performance/accuracy is acceptable or not. In the second case, some action could be taken. This method is in charge to monitor the results received from the uncertainty model. It decides if the current AWM is enough to compute next uncertainty iteration or the AWM and input data (val struct) has to be modified to correspond the new situation at runtime.     
\subsection*{Method onRelease}
Optional, but recommended member function. This environment should eventually contain clean-up of the framework and applications. This is the exit method to execute when the computation finalizes. It is the end of the program and the use must be an user decision. The system eventually does not have to arrive to this point since the system running 24/7  is expected.



\section{Results} 
\label{results}

There are two platforms that serve for development, testing and integration of the example application exposed above. First is a 48 core SMP machine HP ProLiant DL785 G6 \cite{HP14}. A GNU/Linux version 3.13.0-36-generic with a gcc version 4.8.2 is installed on this machine. It consists of 8 sockets with Six-Core AMD Opteron(tm) Processor 8425 HE 64bits 2.1@GHz clock frequency. Each core has a L1 cache memory size of 64KB data and 64KB for instructions and 512KB L2 cache memory and a TLB of 1024 KB. Each CPU socket has an embedded memory 5118 KB L3. The machine is equipped with 256 GB DDR2 533 MHz with ECC memory.  The RTRM system uses resources of the platform, 8 sockets each with six cores in total. One of the cores of each socket, six in total, named hosts, are allocated to monitor the system and the sensors of the platform (i.e. power, temperature etc.) by the RTRM system and to run all non-managed applications and processes. Therefore, there are just 42 real cores for executing managed applications with the RTRM support.

Second platform is HPC cluster Anselm, which consists of 209 compute nodes, totaling 3344 compute cores. Each node is a powerful x86-64 computer, equipped with 16 cores and at least 64GB RAM. Nodes are interconnected by fully non-blocking fat-tree Infiniband network and equipped with Intel Sandy Bridge processors. The cluster runs bullx GNU/Linux operating system, which is compatible with the RedHat GNU/Linux family. There is no RTRM support meantime, but it is part of future work.

After integration of the uncertainty modelling module into the RTRM on SMP machine, we first computed how the coefficient to compute the precision of the simulations decrease with the number of iterations. This coefficient is computed and cumulated in each iteration and heads to the zero at infinity. Lower value of the coefficient means higher precision. These precisions for given numbers of iterations as Fig. \ref{precision} depicts are resources independent and will be used to identify optimal working modes.

\begin{figure}
\centering
\epsfig{file=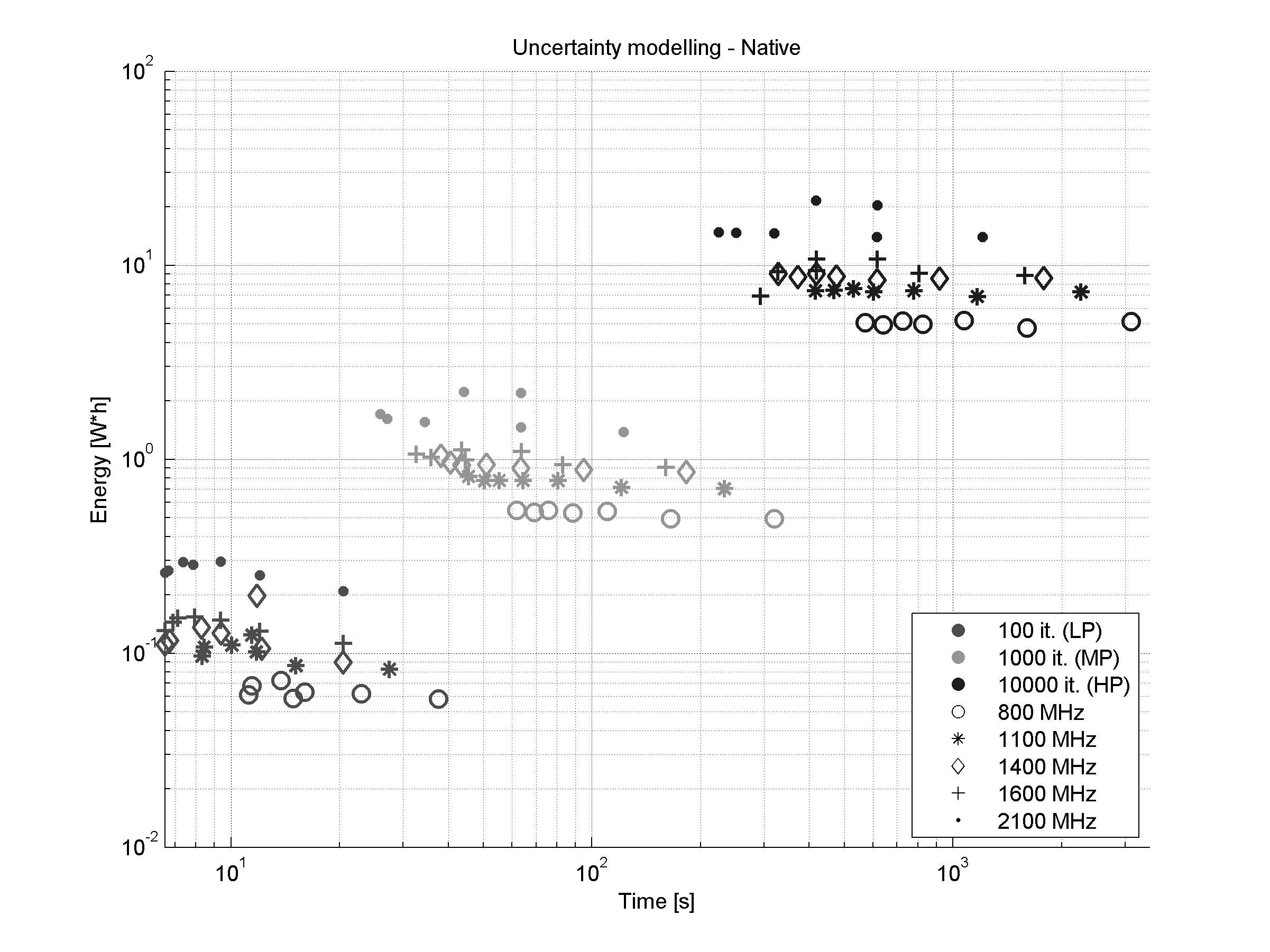, height=2.9in, width=3.8in}
\caption{Uncertainty executed natively, time versus energy consumption, on SMP machine.}
\label{native}
\end{figure}
\begin{figure}
\centering
\epsfig{file=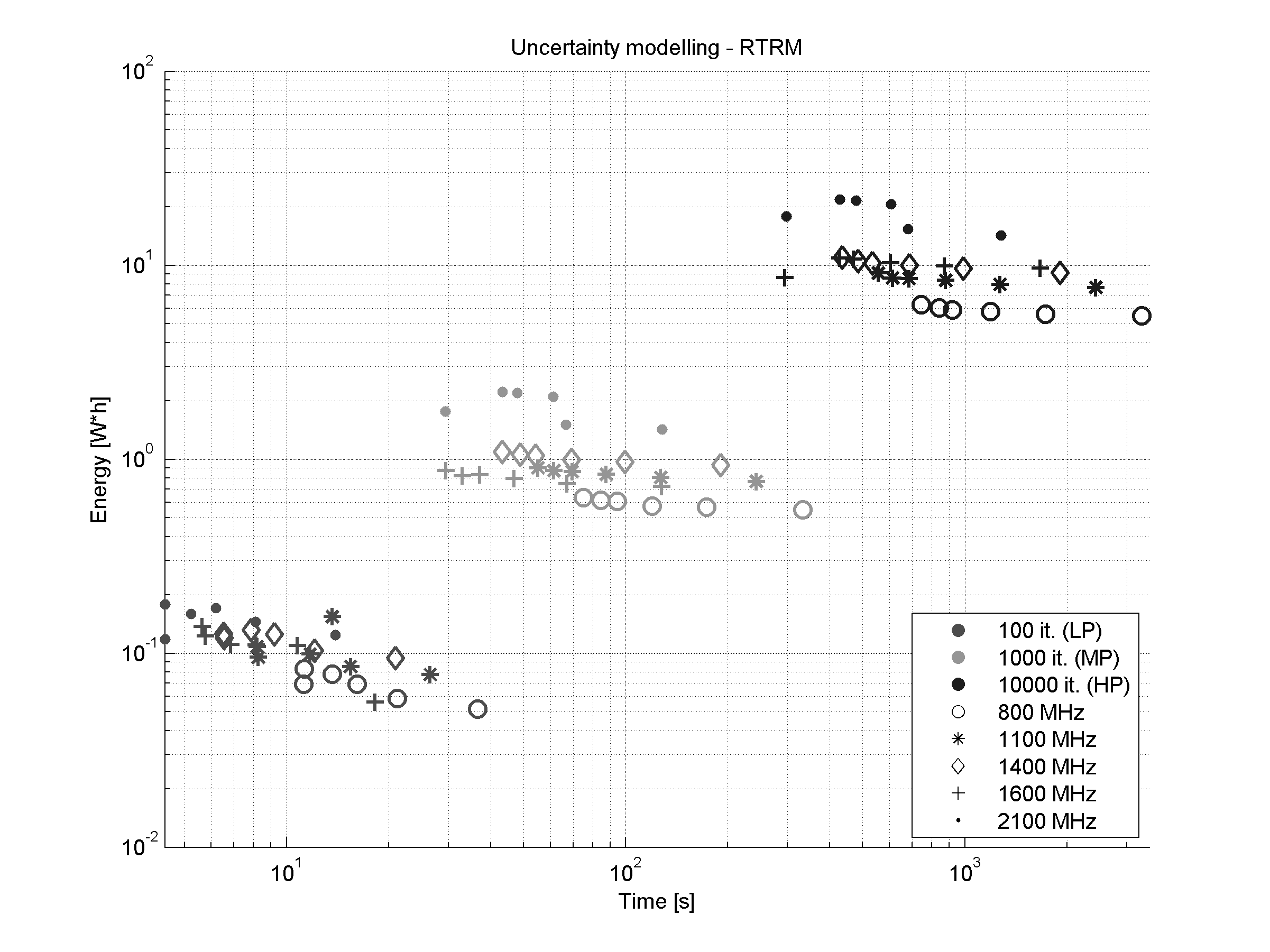,  height=2.9in, width=3.8in}
\caption{Uncertainty executed with RTRM, time versus energy consumption, on SMP machine.}
\label{bbque}
\end{figure}

\begin{figure}
\centering
\epsfig{file=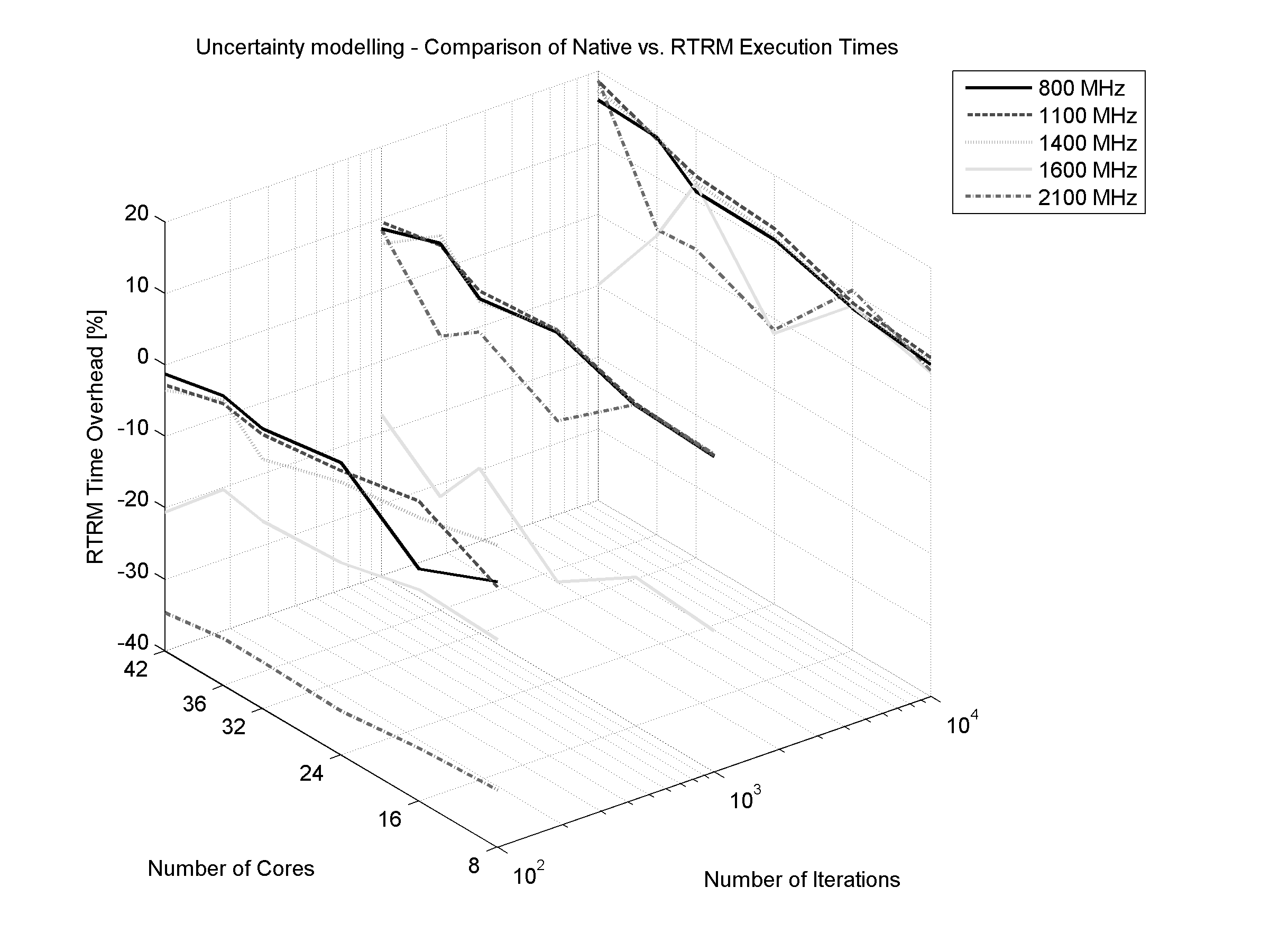,  height=2.9in, width=3.8in}
\caption{Comparison of RTRM time overheads for different configurations.}
\label{diff}
\end{figure}

Fig. \ref{native} presents energy consumption of native, which means unmanaged by RTRM, executions of uncertainty simulations with a galaxy of points. Values are only estimated because of absence of HW support for power consumption measurement of the SMP machine. These points represent simulations with different precision - number of iterations (e.g. 100 low precision, 1000 medium precision or 10000 high precision), different CPU frequencies and various number of cores which results in different execution times and appropriate consumed energy. This leads to trade-offs for running the system with various amount of resources (cores, frequencies).   Fig. \ref{bbque} shows similar results but instead of native execution the simulations run with the RTRM.

Dynamic part (only used cores) of estimated power consumption was computed using original formula:
\begin{equation}P_n = (P_{max} - P_{idle}) \frac{n}{100} +  P_{idle}\end{equation}
Where: 
$n$ is a system load,
$P_n$ is a power consumption at this load,
$P_{max}$ is a maximum power consumption,
$P_{idle}$ is a power consumption in an idle.

Estimated energy consumption $E$ is then computed from the formula:
\begin{equation}E = P  t\end{equation}
Where: $P$ is estimated power consumption and $t$ is time of execution.

\begin{figure}
\centering
\epsfig{file=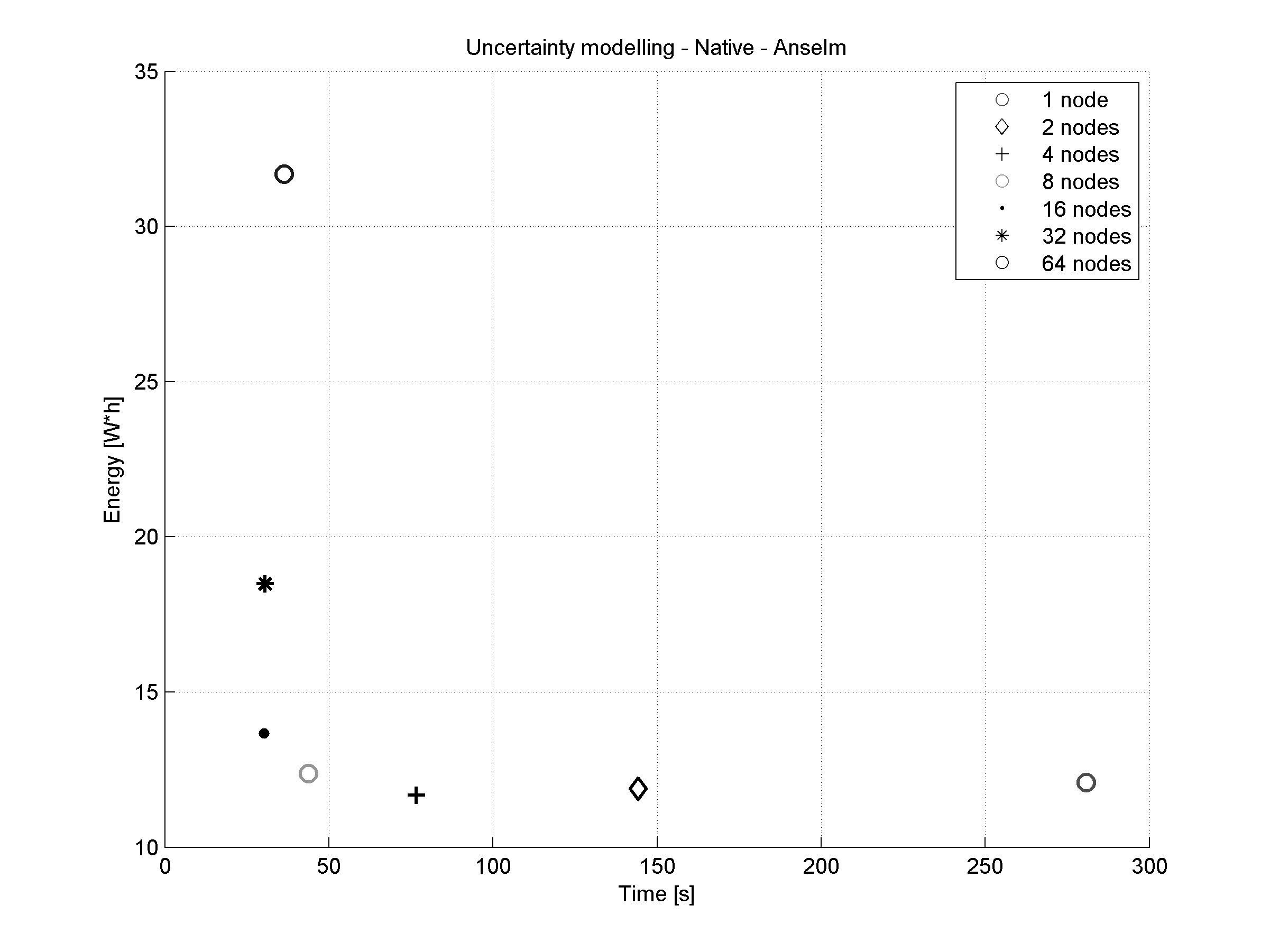,  height=2.9in, width=3.8in}
\caption{Uncertainty executed natively, time versus energy consumption, on HPC cluster.}
\label{hpc}
\end{figure}

We have observed that the overhead of RTRM in time is below 10\% in average, which is shown by Fig. \ref{diff}. We have to take into account that from each socket (six x86-64 cores) of the platform, one core is used to monitor the healthy of the system. So, from 48 possible cores only 42 are enabled to run our uncertainty module. From these points is possible to extract which are the optimal-pareto points that are going to be inserted in the RTMR recipe. For comparison, power consumption of native executions of uncertainty simulations on HPC cluster (multi-node solution using MPI+OpenMP, 10000 iterations) are shown on Fig. \ref{hpc}. Values were measured using the Likwid (likwid-powermeter) tool for different numbers of nodes. The graph shows steeply increasing power consumption with no significant time improvement using more than sixteen 16-cores computing nodes.
 
\section{Conclusion and future work} 

The paper presents an environment that helps to an x86-64 architecture platform to run 24/7  a model saving energy with the trade-off of losing precision with the different resources configurations. The paper shows the uncertainty modelling executed with different steps of precision given by various numbers of iterations leading in appropriate execution times. The decision of precision/executed time/consumed energy is taken by a Run-Time Resource Manager monitoring the workload of the platform and computing the best-trade off with the time overhead less than 10\% in against a native execution. At design time, it is possible to describe the optimal points that are going to be used at runtime. The environment also permits to execute the module 24/7 in nodes of a cluster and/or with accelerators. This enables to use more resources and higher precision for the future.

\section{Acknowledgments}

This article was supported by Operational Programme Education for Competitiveness and co-financed by the European Social Fund within the framework of the project New creative teams in priorities of scientific research, reg. no. CZ.1.07/2.3.00/30.0055, by the European Regional Development Fund in the IT4Innovations Centre of Excellence project (CZ.1.05/1.1.00/02.0070), by the project Large infrastructures for research, development and innovation of Ministry of Education, Youth and Sports of Czech Republic with reg. no. LM2011033, and by 7\textsuperscript{th} EU framework programme project no. FP7-612069 HARPA - Harnessing Performance Variability.

%

\end{document}